\documentclass{article}
\usepackage[dblblindworkshop,final]{neurips_2025}
\usepackage[utf8]{inputenc}
\usepackage[T1]{fontenc}
\usepackage{hyperref}
\usepackage{url}
\usepackage{booktabs}
\usepackage{amsfonts}
\usepackage{nicefrac}
\usepackage{microtype}
\usepackage{xcolor}
\usepackage{multirow}
\usepackage{graphicx}
\usepackage{caption}

\newcommand{\anonymoushospital}{Sankara Eye Hospital}
\newcommand{\cA}{\textit{Write}}
\newcommand{\cB}{\textit{Edit}}
\newcommand{\cC}{\textit{Instruct}}

\title{Editing with AI: How Doctors Refine LLM-Generated Answers to Patient Queries}
\workshoptitle{The Second Workshop on GenAI for Health: Potential, Trust, and Policy Compliance}

\author{
  Rahul Sharma\thanks{Work done while employed at Microsoft Research, Bangalore, India.}\\
  JazzX AI  , Bangalore, India \\
   \And
   Pragnya Ramjee\footnotemark[1]\\
   Stanford University, USA \\
   \And
   Kaushik Murali \\
   Sankara Eye Hospital, Bangalore, India \\
   \texttt{kaushik@sankaraeye.com}
   \And
   Mohit Jain \\
   Microsoft Research, Bangalore, India \\
   \texttt{mohja@microsoft.com} \\
}

\begin{document}
\maketitle

\begin{abstract}
Patients frequently seek information during their medical journeys, but the rising volume of digital patient messages has strained healthcare systems.
Large language models (LLMs) offer promise in generating draft responses for clinicians, yet how physicians refine these drafts remains underexplored.
We present a mixed-methods study with nine ophthalmologists answering 144 cataract surgery questions across three conditions: writing from scratch, directly editing LLM drafts, and instruction-based indirect editing.
Our quantitative and qualitative analyses reveal that while LLM outputs were generally accurate, occasional errors and automation bias revealed the need for human oversight.
Contextualization---adapting generic answers to local practices and patient expectations---emerged as a dominant form of editing.
Editing workflows revealed trade-offs: indirect editing reduced effort but introduced errors, while direct editing ensured precision but with higher workload.
We conclude with design and policy implications for building safe, scalable LLM-assisted clinical communication systems.
\end{abstract}
\section{Introduction}

Patients seek information throughout their medical journeys~\cite{clarke2016healthinfoneeds}.
Effectively addressing these information needs is central to patient-centered care~\cite{zill2015dimensionsofpatientcentredness}, supporting informed decision making~\cite{clarke2016healthinfoneeds} and improving health outcomes like treatment adherence and emotional well-being~\cite{stewart1995effectivephysician-patientcommunication}.
Digital communication platforms, such as chat groups, instant messengers, and web portals, have emerged as powerful tools for meeting these needs, by enabling healthcare providers to deliver timely, accurate responses to patient concerns~\cite{ramjee2025cataractbot,wang2020wechatpatientcarechina,yang2024talk2care}.
However, the increasing ubiquity of these platforms, accelerated by the COVID-19 pandemic and a surge in telemedicine, has strained healthcare systems with unprecedented volumes of interactions~\cite{holmgren2021pandemicehr,north2020providertopatientmessages}.

The advent of generative AI has raised the possibility that large language models (LLMs) could help manage this burden by assisting physicians in patient communication~\cite{ramjee2025cataractbot,sachdeva2024learningslargescaledeploymentllmpowered,yang2024talk2care}.
Early studies evaluating the quality of LLM responses to medical questions found physicians frequently rate these answers as `safe', and sometimes even prefer them to peer-written responses~\cite{singhal2025expertlevelqnawithllms}.
However, LLMs can still output inaccurate, outdated or inappropriate information~\cite{au2023ainotready,biro2025aiinpatientportalmessaging}, and patients report lower trust in invalidated LLM responses~\cite{ramjee2025cataractbot}.
Thus, doctor oversight remains indispensable.

One promising approach is using LLMs to generate \textit{draft} responses to patient queries for clinicians to refine, rather than replace them~\cite{ayers2023physicianandairesponsescomparison, ramjee2025cataractbot}.
Initial deployments report clear benefits such as reduced workload on doctors and improved response quality~\cite{chen2024llmrespondingtopatientlancet,taiseale2024aidraftreplies}.
However, challenges remain, including physicians over-relying on LLM output instead of exercising their own clinical judgment~\cite{biro2025aiinpatientportalmessaging,chen2024llmrespondingtopatientlancet,wadhwa2025designingculturesocialnorms} and producing verbose answers~\cite{taiseale2024aidraftreplies}, potentially due to the labor required in correcting and shortening LLM responses.
Thus, understanding how clinicians refine LLM drafts, and how different co-authoring strategies align with their workflows to maximize usability and effectiveness, is essential~\cite{afshar2024promptengineeringllmresponding,biro2025aiinpatientportalmessaging,garcia2024aigenerateddraftrepliestopatientinbox}.

While the existing body of work has examined LLM response quality and physician acceptance, the specific approaches for refining LLM drafts remain understudied.
We address this gap with a mixed-methods study comparing three answer generation approaches: \textit{writing from scratch} (doctor writes answer, no LLM involved), \textit{direct editing} (manually correcting LLM-generated drafts), and \textit{instruction-based editing} (providing instructions to the LLM for correction).
We evaluate accuracy, completeness, and safety of the answers, along with doctor's efficiency and preferences, to answer these research questions: How do different LLM co-authoring approaches affect physician efficiency and response quality when answering patient queries? What are physicians' perceptions of usability and workflow compatibility across different LLM co-authoring approaches?

To ground our investigation in a concrete clinical context, we focus on cataract surgery, the most common ophthalmic procedure worldwide and the second highest surgical procedure globally~\cite{mcghee2020cataractcommon}.
Doctor-patient communication becomes critical for surgical procedures, as patients seek reassurance about surgical risks and detailed information about post-operative care~\cite{clarke2016healthinfoneeds}.
High patient query volumes have prompted development of LLM-based communication systems, including standardized patient education materials~\cite{thompson2024leafletscataract} and doctor-in-the-loop chatbots~\cite{ramjee2025cataractbot, sachdeva2024learningslargescaledeploymentllmpowered}.

We worked with 9 ophthalmologists who answered 144 cataract surgery questions across the three conditions, then participated in focus group discussions and evaluated answers generated by their peers.
Doctors found LLM-generated answers to be generally accurate and complete, but occasional factual errors and the risk of automation bias underscored the need for human oversight. A central form of editing was contextualization---adapting otherwise generic answers to local practices, terminology, and patient expectations. Editing workflows revealed trade-offs: instruction-based editing reduced effort compared to direct text editing, yet introduced occasional technical errors and ambiguities. Finally, while doctors valued the polished language of LLM drafts, they often had to simplify and reframe answers into shorter, clearer, and more conversational forms suitable for patients.
Together, these results extend prior work by shifting attention from LLM response quality alone to the process of refinement, illuminating design and policy directions for safe, scalable integration of LLMs into patient communication.
\section{Methods}
\label{methods}
We conducted a mixed-method controlled user study during June-July 2025 in collaboration with \anonymoushospital{}, a leading tertiary eye care and teaching hospital in India.
The study took place at its Jaipur and Hyderabad branches.
Approval was obtained from the Scientific and Ethics Committees of both sites.
In line with hospital policy, participants received no financial compensation.

\paragraph{Participants}
Nine practicing ophthalmologists (4 female) from \anonymoushospital{} (4 Hyderabad, 5 Jaipur) participated in the study (Table~\ref{tab:participants}).
All performed at least 10 cataract surgeries per week.
Their mean age was 37.1$\pm$5.5 years, with 11.6$\pm$5.1 years of professional experience.

\begin{figure*}
  \centering
   \includegraphics[width=\textwidth]{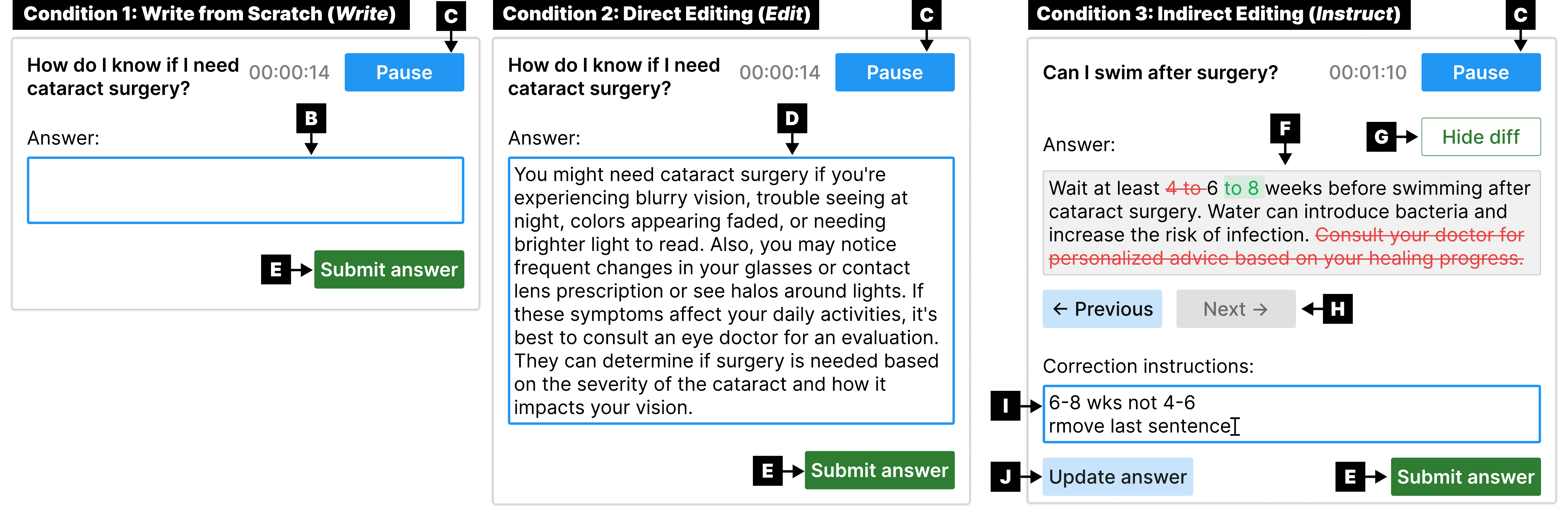} 
    \captionsetup{justification=centering}   
    \caption{Web-based study interface. \textbf{(Left)} Condition 1: \textit{Writing from Scratch} (Write), where doctors write answers in text box \textit{B}. \textbf{(Middle)} Condition 2: \textit{Direct Editing} (Edit), where doctors edit pre-generated LLM answers (in the textbox \textit{D}). \textbf{(Right)} Condition 3: \textit{Indirect Editing} (Instruct), where doctors provide instructions for revision (\textit{I}), with changes visually highlighted (\textit{F}).}
  \label{fig:docedit_portal}
\end{figure*}

\paragraph{Interface Design}
We designed a custom web-based application powered by GPT-4o (Figure~\ref{fig:docedit_portal}) to compare the three conditions.

\textit{Condition 1} (\textit{Writing from Scratch}) aka \cA{}: Doctors answered questions in a blank text box (Figure~\ref{fig:docedit_portal}B, without the pre-generated LLM answer).

\textit{Condition 2} (\textit{Direct Editing}) aka \cB{}: Doctors revised an LLM-generated draft in an editable text box (Figure~\ref{fig:docedit_portal}D).
    
\textit{Condition 3} (\textit{Indirect Editing}) aka \cC{}: Doctors received an LLM draft in a non-editable text box (Figure~\ref{fig:docedit_portal}F) and indirectly edited it by providing instructions in a separate text box (Figure~\ref{fig:docedit_portal}I). The system generated revised responses with changes highlighted (green for additions, red strikethrough for deletions), along with an optional toggle button (Figure~\ref{fig:docedit_portal}G) to hide/show differences. Navigation arrows (Figure~\ref{fig:docedit_portal}H) enabled participant to review past versions of the answer and selecting one for further editing or submission.

All conditions included a timer (Figure~\ref{fig:docedit_portal}C), which doctors could pause only for interruptions.
The order of conditions was randomized across participants for counterbalancing.
All interactions were automatically logged.

\begin{figure*}
  \includegraphics[width=\textwidth]{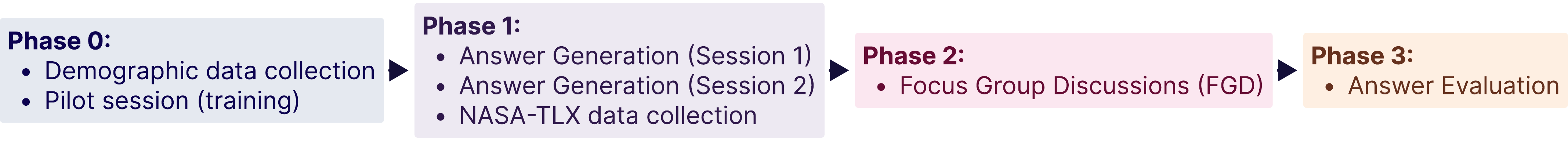}    
    \caption{Phases of our mixed-methods evaluation study.}
  \label{fig:docedit_studydesign}
\end{figure*}

\paragraph{Procedure}
After a pilot study, we conducted the main study in three phases (Figure~\ref{fig:docedit_studydesign}):


\textit{Phase 0: Pilot Study}.
We conducted four training sessions with 9 doctors to familiarize them with the Phase~1 web interface (Figure~\ref{fig:docedit_portal}).
Each session, held on Microsoft Teams and facilitated by two researchers, began with a demonstration of the three conditions using two example questions.
Doctors then practiced with five sample questions per condition. 
We collected demographic information via a web form at the start of each session.
Feedback from these pilots informed minor refinements to the interface.
Each session lasted $\sim$60 minutes.

\textit{Phase 1: Answer Generation}.
We conducted this phase over two sessions with nine doctors, organized into three triplets.
Each doctor generated 48 answers (16 per condition).
The question dataset comprised of 144 common cataract surgery questions sourced from a large-scale patient chatbot deployment in India~\cite{sachdeva2024learningslargescaledeploymentllmpowered}.
In each session, triplets received the same 24 questions, distributed across conditions so that each question was answered once in each condition across the three triplets.
This design enabled inter-condition comparison of response quality and efficiency. 
In total, doctors generated 432 answers.
Afterward, they completed NASA-TLX ratings (5-point Likert scale)~\cite{hancock1988nasatlx} and indicated their most/least preferred, most efficient, safest, and most workflow-aligned condition.

\textit{Phase 2: Focus Group Discussions (FGDs)}.
We conducted three FGD sessions in English via Microsoft Teams, with each doctor participating in one session.
Two researchers facilitated: one moderated, while the other took notes and prompted follow-ups.
To aid recall, screenshots of the different conditions were shared during the discussion. 
Topics included doctors' experiences with each condition, perceived advantages and drawbacks, trust and accuracy of responses, and perspectives on the future of LLM-assisted patient communication.
Each session lasted $\sim$75 minutes and was audio-recorded, with transcripts prepared by a researcher immediately afterward.

\textit{Phase 3: Answer Evaluation}.
Each doctor rated 48 questions (24 from each session), reviewing one answer from each condition per question. 
Ratings covered three dimensions: accuracy, completeness, and likelihood of harm---using the question: `\textit{Please rate the [dimension] of this answer (1 = very low, 2 = low, 3 = medium, 4 = high, 5 = very high).}'
Doctors could optionally provide written justifications.
To minimize bias, all answers were blinded to condition and randomized in order.
Each doctor evaluated answers generated by another triplet to avoid self-assessment.
Each answer received two independent ratings, which we averaged for analysis.

\paragraph{Data Collection and Analysis}
We adopted a mixed-method approach combining quantitative and qualitative analyses. 
Data sources included interaction logs (time, answer length, edit distance), NASA-TLX ratings, preference rankings, FGD transcripts and notes, and answer ratings on accuracy, completeness, and harmfulness. 
Quantitative data (logs and ratings) were analyzed using descriptive statistics, t-tests, and linear mixed-effects models (with condition as a fixed effect and random intercepts for questions and participants). 
Qualitative data (FGD transcripts and notes) were analyzed through inductive thematic analysis~\cite{braun2006thematicanalysis}, initially coded by one author and iteratively discussed with three co-authors to derive broader themes. 
Edits in \cB{} and \cC{} were also categorized and coded.

\begin{table}
\centering
\scriptsize
\caption{Demographic details of study participants (n=9).}
\label{tab:participants}
\begin{tabular}{lllllllll}
\hline
\textbf{ID} &
  \textbf{City} &
  \textbf{Age} &
  \textbf{Gender} &
  \textbf{Prof exp (years)} &
  \textbf{Surgeries/week} &
  \textbf{Participation} \\ \hline
D1 & Hyderabad & 34 & Female & 7 & 10-20 & FGD3\\
D2 & Hyderabad & 36 & Male & 10 & 40+ & FGD1\\
D3 & Hyderabad & 34 & Male & 15 & 40+ & FGD1\\
D4 & Jaipur & 43 & Male & 16 & 20-30 & FGD2\\
D5 & Jaipur & 43 & Male & 17 & 20-30 & FGD2\\
D6 & Jaipur & 45 & Male & 15 & 10-20 & FGD2\\
D7 & Jaipur & 30 & Female & 4 & 20-30 & FGD2\\
D8 & Jaipur & 31 & Female & 5 & 20-30 & FGD2\\
D9 & Hyderabad & 38 & Female & 15 & 40+ & FGD3\\ \hline
\end{tabular}%
\end{table}


\section{Findings}


    
    
    
    
    
    
    

\subsection{Answer Quality and Reliability}

\paragraph{Accuracy, Completeness, and Safety}
Drawings on our interactions with doctors and prior research~\cite{ramjee2025cataractbot}, we focus on three key qualities of doctor-generated answers: accuracy, completeness, and non-harmfulness.
We analyzed 432 answers (3 per question for 144 questions in Phase~1) using ratings from two independent doctors on the three metrics in Phase~3.
Ratings were averaged across the two doctors for analysis.
Linear mixed-model analysis revealed significant differences across conditions (accuracy: $F(2,278)=22.4$, $p<.005$; completeness: $F(2,278)=76.7$, $p<.005$; non-harmfulness: $F(2,278)=23.8$, $p<.005$).
Post-hoc comparisons showed that both \cB{} and \cC{} significantly outperformed \cA{} on all three metrics.
While \cC{} responses were rated slightly higher than \cB{}, these differences were not statistically significant for accuracy ($p=0.07$) or completeness ($p=0.7$). For non-harmfulness, however, \cC{} was rated significantly less harmful than \cB{} ($p<.05$).
Among the 432 answers, harmful ratings were relatively rare: 14 in \cA{}, 7 in \cB{}, and 5 in \cC{}. This suggests that most answers were high-quality and safe for end-users.

Here are examples of low accuracy and high harmfulness (Ans1), and low completeness (Ans2).  
\\
Q1: {\fontfamily{cmss}\selectfont Should I take complete rest post-surgery?}
\\
Ans1: {\fontfamily{cmss}\selectfont Complete bed rest is not recommended. You can resume your daily activities.}
\\
Q2: {\fontfamily{cmss}\selectfont Can the patient take their regular BP medicine on the surgery day?}
\\
Ans2: {\fontfamily{cmss}\selectfont Follow your doctor's instructions carefully to ensure a safe and successful procedure.}

\paragraph{Consistency and Standardization Across Answers}
We observed a philosophical divide among doctors in how answers should be framed.
Some preferred answers tailored to each patient's demographic and context, while others valued standardized responses across patients and doctors.
The first group favored \cA{} as it allowed them to ``\textit{frame answers in their own language},'' whereas the latter leaned toward \cB{} or \cC{} for their consistency.
With an LLM providing a base draft, most doctors made minimal edits, resulting in more homogeneous answers across patients.  
\begin{quote}
    ``\textit{Each doctor in \cA{} will write the answer differently, and also... differently for each patient, as in, if the patient is educated or not... old or young. E.g., I would know a rural patient swims in borewells vs a patient from a city swims in a pool. With that, the answer to a query about swimming becomes location-specific. Manual answers are tailored and very subjective... AI-generated is more homogeneous across geography, across demography.}'' -- D2.
\end{quote}
While some appreciated this uniformity, others criticized LLM-generated answers as ``\textit{generic}'' or ``\textit{lacking depth},'' contrasting with the expectation that doctors be ``\textit{very specific}.''
From a hospital's perspective, however, standardization was seen as beneficial for ensuring uniform care: 
``\textit{If you ask `when to come after the surgery', some surgeons may say next day, some may say next week... The hospital would want a standard treatment guideline, and expect all doctors to say `one week'.}'' -- D5.

\paragraph{Automation Bias as an Emerging Risk}
A consequence of this standardization was the emergence of automation bias.
Doctors reported that they often skipped correcting LLM answers if they were ``\textit{not completely wrong},'' even when the advice diverged from their own practice.
Over time, this tendency risked anchoring doctors to the LLM-generated phrasing and subtly shifting their judgment.  

\begin{quote}
    ``\textit{I edited only where there was strong disagreement. The remaining were okay, so I let it be... E.g., it said `take a head bath after 1 week' while I would say `2 weeks'... I didn't edit this as it's not completely wrong.}'' -- D3.    
\end{quote}

Such omissions increased standardization but reduced subjectivity, creating a risk of clinicians internalizing the AI's defaults.
As D4 reflected: ``\textit{These (LLM-generated) answers can manipulate our thinking. It changes our mindset.}''


\paragraph{Perceived Verbosity vs. Useful Detail}
Answer length varied sharply across conditions. A linear mixed-effects analysis revealed significant differences ($F(2,278)=536.9$, $p<.001$): \cA{} produced much shorter answers (89.6$\pm$74.5 characters/answer) than both \cB{} (335.1$\pm$90.1) and \cC{} (351.6$\pm$103.8), with both pairwise comparisons showing p<.001. No difference was found between \cB{} and \cC{} (p=0.7).
Comparing the two edit conditions, \cB{} and \cC{}, we found no statistical differences in insertion, deletion, substitution, or overall edit distance.

Doctors' editing behavior highlighted how length shaped their experience.
In \cB{}, only 34 of 144 answers (23.6\%) were edited.
Edits typically shortened answers, reducing their length from 355.7$\pm$68.9 to 264.1$\pm$127.3 characters.
Edits included deleting unnecessary details (e.g., D2 removed: ``{\fontfamily{cmss}\selectfont Also, medications used during surgery might cause drowsiness.}'') or condensing verbose phrasing.
Additions of new content were rare (14 answers) compared to deletions (39) and substitutions (27).

In \cC{}, editing unfolded differently. Of the 144 answers, 51 (35.4\%) were edited, requiring 134 instructions in total (2.6$\pm$1.4 per answer). Two answers required as many as seven iterative comments.
Doctors noted the difficulty of condensing LLM-generated answers through instructions:  

\begin{quote}
    ``\textit{Every AI-generated answer usually is lengthy... and has multiple points. For each point, I need to do multiple modifications... I can't give instructions at one go. I have to keep modifying.}'' -- D8
\end{quote}

Taken together, these results show that while LLM-supported conditions (\cB{}, \cC{}) generated more detailed and comprehensive answers than \cA{}, doctors often worked to shorten or streamline them.
This tension suggests that answer length, though positively associated with completeness, could burden clinicians with extra review effort or risk overwhelming patients with information.

\subsection{Editing Workflows and Doctor Effort}

\paragraph{Cognitive and Physical Demands}
NASA-TLX ratings revealed clear differences across conditions:
\cC{} was rated as the least physically and mentally demanding, while \cA{} was rated the highest on both dimensions.
Doctors stated that \cA{} felt ``\textit{cumbersome}'' (D9) because it required extensive typing and recall to compose answers from scratch.
In contrast, \cB{} and \cC{} shifted effort toward reviewing and editing existing drafts.
Editing in \cB{} often involved small textual adjustments, while \cC{} enabled concise, instruction-based edits: ``\textit{I provided very short, crisp instructions.}'' (D5).

The nature of mental effort varied by condition.
In \cA{}, doctors worried about completeness: ``\textit{I always thought `}Am I forgetting something?\textit{'... E.g., for the Q on `do's and don'ts for cataract surgery', that's a long list. It is hard to get everything.}'' (D9).
In \cB{}, effort centered on scanning long drafts for issues: ``\textit{There is a maze of information in the AI answers... It's not trivial to read 8 lines and decide what to add, where to add it, and what to delete.}'' (D5).
In \cC{}, cognitive load came from sequencing effective instructions and verifying results: ``\textit{I need to think what is there and not, what instruction needs to be given... and whether that instruction worked or not.}'' (D1).
Some doctors found this cognitive model appealing---``\textit{computer making the edit is better, as my thoughts fit well with it}'' (D2)---while others pointed the learning curve and stochasticity involved in \cC{}: ``\textit{What command to give, how to make it work... there is also the uncertainty of how my instruction will be interpreted}'' (D9).

Editing efficiency reflected these trade-offs.
A linear mixed-effects analysis revealed significant differences across conditions ($F(2,278)=41.3$, $p<.001$).
\cA{} (52.0$\pm$44.5s per answer) took significantly longer than both \cB{} (27.9$\pm$39.9s) and \cC{} (46.4$\pm$54.3s) (p<.001), and \cC{} was also slower than \cB{} (p<.05).
This pattern is partly explained by the fact that only a fraction of answers were edited in \cB{} (23.6\%) and \cC{} (35.4\%).
However, when restricting analysis to edited answers, \cC{} (85.3$\pm$60.4s) was significantly slower than \cA{} (52.0$\pm$44.4s), while \cB{} (50.2$\pm$37.2s) did not differ from either.
Focus group discussions suggested this was partly due to the novelty of \cC{}, which required time to formulate instructions, review edits, and manage uncertainty about corrections.
Although doctors noted a small wait time for the LLM to process instructions in \cC{}, it was not statistically significant.

\paragraph{Types of Edits}
Manual analysis of the 84 corrections in \cB{} and \cC{} revealed four categories: corrections to incorrect or ambiguous content (27), deletions of redundant details (63), additions of missing or context-specific information (43), and rewording or restructuring for clarity~(9).
Each reflected distinct limitations of the LLM drafts and clinicians' priorities in tailoring them for patients.


\textit{Corrections}.
Several answers required factual correction, often because the LLM produced misleading or overly generalized statements.
For example, to the question ``{\fontfamily{cmss}\selectfont Do they put a needle in your eye during cataract surgery?}'', the LLM-generated answer was ``{\fontfamily{cmss}\selectfont No, a needle is not put into your eye during cataract surgery. The surgery is...}'', which D4 corrected in \cB{} to ``{\fontfamily{cmss}\selectfont Yes, needles are used in a standard way for some steps in cataract surgery. The surgery is...}''.
For the same question, D5 in \cC{} simply instructed the system to ``{\fontfamily{cmss}\selectfont remove the first sentence}''.
Similarly, D6 flagged that the LLM's advice on fasting before surgery---``{\fontfamily{cmss}\selectfont no food for six hours}''---was inaccurate, since cataract surgery is typically performed under local anesthesia, requiring only two hours of fasting, with a light breakfast permitted.
Beyond factual inaccuracies, doctors also corrected vague or hedged responses.
As D5 put it, ``\textit{Patients want concrete, clear answers. AI is overly cautious and ambivalent. For a question like `Can I swim?', the answer should be definitive... `no, not for the first month.'}''

\textit{Deletions}.
All doctors shortened answers by removing verbose or repetitive content.
In total, 63 deletions were recorded, including 34 single-sentence removals, 20 two-sentence deletions, and 9 cases of deleting more than two sentences.
Doctors felt that long, elaborate answers risked confusing patients, who typically preferred concise advice. As D4 emphasized, ``\textit{Shorter, to-the-point answers are definitely better.}''
D5 elaborated: ``\textit{With AI, a short answer becomes a long answer. E.g., `Can I swim after surgery?' I would simply say `no'. But AI will say the same `no' in a more appropriate, but very lengthy manner... In face-to-face conversations, I would have given a short `no' as answer.}''
Doctors also ensured that the final answer contained as much ``\textit{direct information}'' as possible.
Redundant closing statements such as ``consult your doctor'' were almost universally removed.

\textit{Additions and Rewording.}
Doctors also added clarifying details that the LLM drafts had omitted, ensuring responses were complete and clinically useful.
In total, 43 such additions were recorded, often involving patient-relevant advice.
For example, D9 added practical lifestyle advice such as ``{\fontfamily{cmss}\selectfont Avoid sunlight and dusty environments}'' as post-surgery precautions, which the LLM had omitted.
Rewording was less frequent (9 cases), but served to improve readability and presentation.
Doctors sometimes requested alternative formats such as bullet points, or relied on \cC{}'s ability to automatically adjust grammar and flow.

Across categories, edits reflected a shared goal: make responses direct, clear, and patient-appropriate.


\paragraph{Contextualization}
Beyond these structural edits, doctors frequently engaged in contextualization. 
Although the study was conducted in India, the prompt used to generate the initial LLM answers was not tailored to this context.
As a result, the answers in \cB{} and \cC{} were often generic and universal in tone, lacking important cultural, clinical, and geographic grounding. 
Doctors frequently intervened to localize these answers so they would be relevant and actionable for their patients. 
Typical edits included substituting technical terms (e.g., replacing ``\textit{ECCE}'' with the more commonly used ``\textit{SICS}'' surgery type), correcting currency references from USD to  INR when discussing costs, and adding India-specific lifestyle guidance such as ``{\fontfamily{cmss}\selectfont avoid sunlight and dusty environments}.'' 
Contextualization spanned \emph{all} edit types---corrections, additions, deletions, and rewording.
Clinicians emphasized that local adaptation was essential for patient comprehension and trust, and that answers without such contextualization risked feeling disconnected from the realities of Indian patients.
At the same time, a few admitted that they sometimes skipped minor contextual corrections when the AI's response appeared ``\textit{good enough},'' echoing the broader risk of automation bias.

\paragraph{Editing Strategies and Challenges}
When faced with major disagreements with the LLM-generated drafts, doctors adopted very different strategies in \cB{} and \cC{}.  
In \cB{}, the common approach was to delete the entire draft and effectively revert to \cA{}, composing a fresh answer from scratch.
In contrast, in \cC{}, doctors often either provided instructions to regenerate the response or rewrote the full answer within the instruction itself. As D9 explained: ``\textit{If it is a very big answer, and I don’t agree with it (in \cC{}), I need to write out the full answer as instructions.}''
In \cC{}, we observed two distinct strategies: issuing multiple short sequential instructions each fixing a single issue, or combining several fixes into one longer instruction. For example, one doctor combined several fixes into a single instruction:
``{\fontfamily{cmss}\selectfont Remove line 1. Add `It is advisable to avoid cooking in the first week after cataract surgery.' Line 2: `However, if required, cooking can be done, but it's important to be careful.' Remove line 3. Remove line 6.}''  

With respect to smaller edits, doctors reported two main challenges in \cB{}: difficulty integrating new content seamlessly into the draft, and the need to repeat similar edits across multiple sections.
For integration, doctors noted that adding a new sentence often disrupted the flow and required careful matching of grammar and style. As D2 explained: 
\begin{quote}
    ``\textit{I don't like \cB{}... A newly added sentence needs to be of the same grammatical standard as the AI answer. It is hard to figure out where to fit this new sentence. It takes time and thinking.}'' -- D2.
\end{quote}
In contrast, the same participant found \cC{} far easier: ``\textit{Half the job is done. Adding anything here is much easier. It takes care of grammar, where to add the new points... framing, everything!}'' -- D2.
A second issue was redundancy.
Because similar phrasing could recur multiple times, doctors had to manually correct each instance.
D5 described: ``\textit{In \cB{}, I might need to edit the first part of a sentence, but similar content may recur again in the next sentence, which I have to edit again. Such inconsistencies go away in \cC{} automatically.}''  
In our manual review of \cB{} outputs, we found 9 answers (across 6 doctors) containing informal phrasing, punctuation errors, and grammatical mistakes--problems that were not present in \cC{} due to its automated re-framing of edits.  

\paragraph{Smartphone-Based Editing Constraints}
Although doctors were instructed to use a laptop/desktop for Phase~1--since the interface was not optimized for small screens--three doctors completed one session on their smartphones.
As D5 explained: ``\textit{The phone is always with me, I prefer that. I rarely sit at a computer... In a real-world scenario, there is a higher probability that I use my phone for such tasks.}''
All three doctors unanimously preferred \cC{} over \cB{} on smartphones, stating that direct text edits were cumbersome.
They noted that tasks such as correcting a spelling required ``\textit{moving the cursor and clicking before editing}'' (D3), and that deleting long text meant ``\textit{holding down the back button for a long time}'' (D4)---both difficult on a phone.

\subsection{Role and Limitations of LLMs}

\paragraph{LLM-Generated Answers Being Lengthy, Inaccurate, and Formal}
Doctors generally described the LLM-generated answers in \cB{} and \cC{} as ``\textit{well-written}'', ``\textit{complete}'', and ``\textit{accurate}''. 
As D2 remarked, ``\textit{AI didn't make any blunders. I edited a few timelines... but these are specific to each doctor,}'' while D6 noted, ``\textit{Most of the (LLM) answers were correct. Only a few minor edits were required.}'' 
At the same time, doctors consistently highlighted three limitations.  
First, answers were often unnecessarily long and verbose: what could be conveyed in one or two sentences was typically expanded into a multi-paragraph explanation, which doctors felt risked confusing patients.  
Second, the content occasionally contained factual inaccuracies or generic statements not aligned with standard cataract practice in India, e.g., instructions about fasting appropriate for general anesthesia but not for local anesthesia. Such errors required careful correction.  
Finally, the tone of the answers was described as overly formal and academic, often framed in textbook-like language.
While doctors appreciated the grammatical polish, they emphasized that such formality could feel impersonal and disconnected from patient communication.
In their edits, they sought to make responses more direct, conversational, and reassuring.  
As D8 summarized for \cB{}: ``\textit{The AI answer requires only minimal modifications, but I still change it to sound more natural for the patient.}''

\paragraph{Technical Errors in \cC{}}
Manual analysis and focus group discussions revealed five instances where \cC{} failed to execute editing instructions exactly as doctors intended. 
E.g., D2 noted that asking the system to ``{\fontfamily{cmss}\selectfont remove the third sentence}'' instead deleted the second, while D1 reported that ``{\fontfamily{cmss}\selectfont remove from eating onwards}'' removed two additional sentences. 
In such cases, doctors adopted two strategies: re-issuing the instruction more explicitly, or using recovery options such as the ``Previous'' button or typing ``{\fontfamily{cmss}\selectfont undo}'' as the next instruction.
These worked because the system maintained conversation history.
D5 recalled: ``\textit{I asked it to remove the second and fourth sentences. However, it only removed the second. So I had to ask again to remove the third, and it worked.}''
These safeguards, along with the ability to see highlighted changes, helped doctors catch and correct errors quickly, though they remained cautious about how their instructions would be interpreted.

\section{Discussion}
\label{discussion}

\paragraph{Keeping Human Experts-in-the-Loop}
Our findings highlight the need to maintain a strong safeguard by keeping doctors firmly in the verification loop for all LLM-generated answers.
While LLMs often produced drafts that were accurate and complete, occasional inaccuracies, omissions, or overly generic advice highlighted the risks of unchecked automation, consistent with prior work~\cite{au2023ainotready,biro2025aiinpatientportalmessaging,ramjee2025cataractbot}.
In the current medicolegal landscape, accountability for medical guidance ultimately rests with licensed practitioners, not AI systems.
This makes human review indispensable, both for patient safety and for ensuring legitimacy of the information provided.

At the same time, there is a practical challenge of scalability.
If every patient poses numerous questions, doctors cannot feasibly review all responses without significant burden.
This opens a design and workforce question: who should validate LLM outputs at scale?
One possibility is task-shifting, where trained non-doctor professionals (e.g., nurse educators, patient counselors) provide frontline validation of LLM answers, escalating only high-risk or ambiguous cases to physicians.
Such tiered review mechanisms would align oversight intensity with the stakes of the advice, preserving safety without overwhelming experts.
Their feasibility is evident in prior ethnographic work, where nurses in resource-constrained settings served as the first line of response to patient queries in digital chat groups~\cite{wang2020wechatpatientcarechina}, albeit without LLMs involved.

Policy frameworks will also need to clarify accountability and medicolegal responsibility for LLM-generated patient information.
Clear guardrails on liability can both protect patients and provide confidence for clinicians to engage with these systems.

\paragraph{Balancing Standardization and Subjectivity}
A central tension in our findings lies between the standardization offered by LLMs and the subjective nuances contributed by individual doctors.
On the positive side, AI-mediated standardization reduces variability in patient education---ensuring that every patient receives consistent and reliable core information.
This consistency may reduce confusion and inequities in care that can arise from idiosyncratic differences across clinicians.
Yet this consistency comes at a cost.
The flip side of standardization is a loss of personalization.
Patients may miss out on the individualized communication styles and nuanced emphases that human doctors naturally bring to their explanations.
Over time, over-standardization risks making patient education less empathetic, less culturally attuned, and less responsive to individual concerns.

A further risk is overreliance.
Because most LLM-generated answers were found to be ``good enough,'' there is a temptation for clinicians to skip detailed review--a risk highlighted in prior work as well~\cite{biro2025aiinpatientportalmessaging, chen2024llmrespondingtopatientlancet, wadhwa2025designingculturesocialnorms}.
This echoes broader concerns in medical AI around automation bias: clinicians may gradually outsource judgment to AI systems, potentially eroding their clinical acumen and vigilance.
Guardrails, such as requiring justification of edits, introducing periodic audits, or leveraging crowdsourced expert edits~\cite{ramjee2025ashabot} may help mitigate this drift while preserving the efficiency of LLM-generated drafts.
Design implications flow directly from these tensions~\cite{inform-bot-design-dis18}.
Interfaces should foreground contextualization---making it easier for doctors to tailor generic LLM drafts to local cultural, clinical, and geographic realities.
Systems should also embed transparency mechanisms, such as highlighting changes, to sustain clinician engagement and reduce the risk of blind acceptance.
Finally, to preserve sustainability, tools should help minimize redundant edits and enable efficient review workflows, so that the efficiency gains of AI do not come at the cost of clinician burnout.

\paragraph{Design Implications}

\textit{Qualities of Effective Answers}.
Our findings highlight that useful answers share core qualities: they should be short, direct, and free of unnecessary complexity.
Patients expect clarity and commitment rather than hedging or vague phrasing.
For example, effective responses clearly state what is and is not allowed, often with timelines, instead of offering conditional or overly cautious language.
This aligns with prior work on patient-centered communication~\cite{chen2025healthtechforimmigrants,liu2025humanizedllms}, which emphasizes plain language and actionable guidance.
Embedding these principles into LLM prompting and fine-tuning could improve default outputs and reduce the need for rewording by doctors.

\textit{Improving LLM-Generated Outputs}.
While LLMs produced fluent drafts, they often lacked contextual grounding.
Recommendations sometimes referred to procedures or practices not followed in India, used inappropriate currency or units, or generated verbose multi-paragraph explanations.
To reduce such mismatches, prompts should explicitly incorporate geographic and institutional context (similar to \cite{ramjee2025ashabot}), and default outputs should be optimized for brevity and clarity.
A design strategy could be to generate short, two-sentence responses by default, with optional expansion for additional detail.
Such changes would reduce editing burden and help ensure that answers remain both relevant and comprehensible to patients.

\textit{Supporting Hybrid Editing Workflows}.
Editing strategies differed across conditions: \cB{} was efficient for small corrections but cumbersome for major rewrites, while \cC{} handled grammar and flow well but was less effective for deletions or large changes.
To accommodate this diversity, future systems should support hybrid workflows.
For example, \cA{} could be paired with LLM-based grammar refinement to produce polished drafts from doctor-authored text.
\cB{} could integrate options for deleting large sections or starting fresh, while \cC{} should include a transparent undo button to recover from misinterpretations.
Designing for fluid transitions between modes can reduce cognitive load and support seamless editing, consistent with principles of mixed-initiative interaction~\cite{amershi2014powertothepeople,horvitz1999mixedinitiative} and recent work on AI-assisted writing~\cite{reza2024abscribehumanaicowriting}.

\textit{Voice Input for Mobile Use}.
Because many clinicians accessed the system via smartphones, typing was described as slow and screen-cluttering.
Voice input emerged as a natural modality, particularly for \cA{} and \cC{}, where answers are short or instructions conversational.
However, voice is less compatible with cursor-based direct editing in \cB{}.
A promising direction is to combine voice dictation for drafting with lightweight correction tools for transcription errors.
This approach builds on prior HCI findings~\cite{ramjee2025ashabot, ramjee2025cataractbot} that multimodal input, especially voice, can reduce physical effort and improve efficiency in mobile, resource-constrained settings.

\textit{Personalizing to Patient Values}.
Finally, answers should adapt not only to clinical accuracy but also to patient needs and preferences.
Older patients may benefit from concise, directive responses, while more educated or information-seeking patients may prefer detailed explanations.
Similarly, some patients require reassurance and empathy, while others prioritize brevity.
This calls for adaptive answer generation that adjusts length, tone, and level of detail based on patient demographics and values.
Prior work~\cite{chen2025healthtechforimmigrants,joshi2025userpereferencesaihealth} in health communication demonstrates that such tailoring improves comprehension, trust, and adherence, suggesting a valuable direction for LLM-assisted systems.

\paragraph{Limitations}\label{limitations}
Although we made specific attempts to reduce the skew in our participant cohort, we used convenience sampling in our recruitment, which may introduce selection bias and affect the representativeness of the sample. Further, we worked with a population in urban India, and views of chatbots and their advice may be different in rural or periurban contexts.
Finally, while doctors were central evaluators, patients--the ultimate consumers of these answers--were not included, and their perspectives remain an important direction for future work.

\paragraph{Conclusion}
This paper examined how doctors answered patient queries using three approaches: writing from scratch, directly editing LLM drafts, and instruction-based indirect editing. While LLMs generated accurate and polished responses, doctors' edits highlighted essential needs: contextualizing content to local practices, safeguarding against factual errors, and reframing verbose responses into clear, patient-centered communication.
Our findings emphasize the importance of keeping human experts in the loop, not only for safety, but also to sustain personalization and empathy in patient education.
This work contributes empirical evidence on human–AI co-authoring in high-stakes settings and identifies design opportunities.
Ultimately, safe and scalable adoption of LLMs in healthcare will depend on balancing standardization with subjectivity, and automation with meaningful human oversight.

\bibliographystyle{plain}
\bibliography{references}

@Article{chen2024llmrespondingtopatientlancet,
author={Chen, Shan
and Guevara, Marco
and Moningi, Shalini
and Hoebers, Frank
and Elhalawani, Hesham
and Kann, Benjamin H.
and Chipidza, Fallon E.
and Leeman, Jonathan
and Aerts, Hugo J. W. L.
and Miller, Timothy
and Savova, Guergana K.
and Gallifant, Jack
and Celi, Leo A.
and Mak, Raymond H.
and Lustberg, Maryam
and Afshar, Majid
and Bitterman, Danielle S.},
title={The effect of using a large language model to respond to patient messages},
journal={The Lancet Digital Health},
year={2024},
month={Jun},
day={01},
publisher={Elsevier},
volume={6},
number={6},
pages={e379-e381},
issn={2589-7500},
doi={10.1016/S2589-7500(24)00060-8},
url={https://doi.org/10.1016/S2589-7500(24)00060-8}
}

@Article{biro2025aiinpatientportalmessaging,
author={Biro, Joshua M.
and Handley, Jessica L.
and Malcolm McCurry, J.
and Visconti, Adam
and Weinfeld, Jeffrey
and Gregory Trafton, J.
and Ratwani, Raj M.},
title={Opportunities and risks of artificial intelligence in patient portal messaging in primary care},
journal={npj Digital Medicine},
year={2025},
month={Apr},
day={24},
volume={8},
number={1},
pages={222},
issn={2398-6352},
doi={10.1038/s41746-025-01586-2},
url={https://doi.org/10.1038/s41746-025-01586-2}
}

@article{taiseale2024aidraftreplies,
    author = {Tai-Seale, Ming and Baxter, Sally L. and Vaida, Florin and Walker, Amanda and Sitapati, Amy M. and Osborne, Chad and Diaz, Joseph and Desai, Nimit and Webb, Sophie and Polston, Gregory and Helsten, Teresa and Gross, Erin and Thackaberry, Jessica and Mandvi, Ammar and Lillie, Dustin and Li, Steve and Gin, Geneen and Achar, Suraj and Hofflich, Heather and Sharp, Christopher and Millen, Marlene and Longhurst, Christopher A.},
    title = {AI-Generated Draft Replies Integrated Into Health Records and Physicians’ Electronic Communication},
    journal = {JAMA Network Open},
    volume = {7},
    number = {4},
    pages = {e246565-e246565},
    year = {2024},
    month = {04},
    issn = {2574-3805},
    doi = {10.1001/jamanetworkopen.2024.6565},
    url = {https://doi.org/10.1001/jamanetworkopen.2024.6565},
}

@ARTICLE{afshar2024promptengineeringllmresponding,
  title    = "Prompt engineering with a large language model to assist
              providers in responding to patient inquiries: a real-time
              implementation in the electronic health record",
  author   = "Afshar, Majid and Gao, Yanjun and Wills, Graham and Wang, Jason
              and Churpek, Matthew M and Westenberger, Christa J and Kunstman,
              David T and Gordon, Joel E and Goswami, Cherodeep and Liao, Frank
              J and Patterson, Brian",
  journal  = "JAMIA Open",
  volume   =  7,
  number   =  3,
  pages    = "ooae080",
  month    =  aug,
  year     =  2024,
  address  = "United States",
  language = "en"
}

@article{garcia2024aigenerateddraftrepliestopatientinbox,
    author = {Garcia, Patricia and Ma, Stephen P. and Shah, Shreya and Smith, Margaret and Jeong, Yejin and Devon-Sand, Anna and Tai-Seale, Ming and Takazawa, Kevin and Clutter, Danyelle and Vogt, Kyle and Lugtu, Carlene and Rojo, Matthew and Lin, Steven and Shanafelt, Tait and Pfeffer, Michael A. and Sharp, Christopher},
    title = {Artificial Intelligence–Generated Draft Replies to Patient Inbox Messages},
    journal = {JAMA Network Open},
    volume = {7},
    number = {3},
    pages = {e243201-e243201},
    year = {2024},
    month = {03},
    issn = {2574-3805},
    doi = {10.1001/jamanetworkopen.2024.3201},
    url = {https://doi.org/10.1001/jamanetworkopen.2024.3201},
}

@article{zill2015dimensionsofpatientcentredness,
  title={Which dimensions of patient-centeredness matter?-Results of a web-based expert delphi survey},
  author={Zill, J{\"o}rdis M and Scholl, Isabelle and H{\"a}rter, Martin and Dirmaier, J{\"o}rg},
  journal={PloS one},
  volume={10},
  number={11},
  pages={e0141978},
  year={2015},
  publisher={Public Library of Science San Francisco, CA USA}
}

@article{clarke2016healthinfoneeds,
author = {Martina A Clarke and Joi L Moore and Linsey M Steege and Richelle J Koopman and Jeffery L Belden and Shannon M Canfield and Susan E Meadows and Susan G Elliott and Min Soon Kim},
title ={Health information needs, sources, and barriers of primary care patients to achieve patient-centered care: A literature review},

journal = {Health Informatics Journal},
volume = {22},
number = {4},
pages = {992-1016},
year = {2016},
doi = {10.1177/1460458215602939},
    note ={PMID: 26377952},
URL = { https://doi.org/10.1177/1460458215602939 },
eprint = { https://doi.org/10.1177/1460458215602939 }
}

@ARTICLE{stewart1995effectivephysician-patientcommunication,
  title    = "Effective physician-patient communication and health outcomes: a
              review",
  author   = "Stewart, M A",
  journal  = "CMAJ",
  volume   =  152,
  number   =  9,
  pages    = "1423--1433",
  month    =  may,
  year     =  1995,
  address  = "Canada",
  language = "en"
}

@ARTICLE{north2020providertopatientmessages,
  title    = "A Retrospective Analysis of {Provider-to-Patient} Secure
              Messages: How Much Are They Increasing, Who Is Doing the Work,
              and Is the Work Happening After Hours?",
  author   = "North, Frederick and Luhman, Kristine E and Mallmann, Eric A and
              Mallmann, Toby J and Tulledge-Scheitel, Sidna M and North, Emily
              J and Pecina, Jennifer L",
  journal  = "JMIR Med Inform",
  volume   =  8,
  number   =  7,
  pages    = "e16521",
  month    =  jul,
  year     =  2020,
  address  = "Canada",
  language = "en"
}

@article{holmgren2021pandemicehr,
    author = {Holmgren, A Jay and Downing, N Lance and Tang, Mitchell and Sharp, Christopher and Longhurst, Christopher and Huckman, Robert S},
    title = {Assessing the impact of the COVID-19 pandemic on clinician ambulatory electronic health record use},
    journal = {Journal of the American Medical Informatics Association},
    volume = {29},
    number = {3},
    pages = {453-460},
    year = {2021},
    month = {12},
    issn = {1527-974X},
    doi = {10.1093/jamia/ocab268},
    url = {https://doi.org/10.1093/jamia/ocab268},
    eprint = {https://academic.oup.com/jamia/article-pdf/29/3/453/42333275/ocab268.pdf},
}

@article{ramjee2025cataractbot,
author = {Ramjee, Pragnya and Sachdeva, Bhuvan and Golechha, Satvik and Kulkarni, Shreyas and Fulari, Geeta and Murali, Kaushik and Jain, Mohit},
title = {CataractBot: An LLM-powered Expert-in-the-Loop Chatbot for Cataract Patients},
year = {2025},
issue_date = {June 2025},
publisher = {ACM},
address = {New York, NY, USA},
volume = {9},
number = {2},
url = {https://doi.org/10.1145/3729479},
doi = {10.1145/3729479},
journal = {Proc. ACM Interact. Mob. Wearable Ubiquitous Technol.},
month = jun,
articleno = {45},
numpages = {31},
}

@article{yang2024talk2care,
author = {Yang, Ziqi and Xu, Xuhai and Yao, Bingsheng and Rogers, Ethan and Zhang, Shao and Intille, Stephen and Shara, Nawar and Gao, Guodong Gordon and Wang, Dakuo},
title = {Talk2Care: An LLM-based Voice Assistant for Communication between Healthcare Providers and Older Adults},
year = {2024},
issue_date = {June 2024},
publisher = {Association for Computing Machinery},
address = {New York, NY, USA},
volume = {8},
number = {2},
url = {https://doi.org/10.1145/3659625},
doi = {10.1145/3659625},
journal = {Proc. ACM Interact. Mob. Wearable Ubiquitous Technol.},
month = may,
articleno = {73},
numpages = {35}
}

@Article{singhal2025expertlevelqnawithllms,
author={Singhal, Karan
and Tu, Tao
and Gottweis, Juraj
and Sayres, Rory
and Wulczyn, Ellery
and Amin, Mohamed
and Hou, Le
and Clark, Kevin
and Pfohl, Stephen R.
and Cole-Lewis, Heather
and Neal, Darlene
and Rashid, Qazi Mamunur
and Schaekermann, Mike
and Wang, Amy
and Dash, Dev
and Chen, Jonathan H.
and Shah, Nigam H.
and Lachgar, Sami
and Mansfield, Philip Andrew
and Prakash, Sushant
and Green, Bradley
and Dominowska, Ewa
and Ag{\"u}era y Arcas, Blaise
and Toma{\v{s}}ev, Nenad
and Liu, Yun
and Wong, Renee
and Semturs, Christopher
and Mahdavi, S. Sara
and Barral, Joelle K.
and Webster, Dale R.
and Corrado, Greg S.
and Matias, Yossi
and Azizi, Shekoofeh
and Karthikesalingam, Alan
and Natarajan, Vivek},
title={Toward expert-level medical question answering with large language models},
journal={Nature Medicine},
year={2025},
month={Mar},
day={01},
volume={31},
number={3},
pages={943-950},
issn={1546-170X},
doi={10.1038/s41591-024-03423-7},
url={https://doi.org/10.1038/s41591-024-03423-7}
}

@ARTICLE{ayers2023physicianandairesponsescomparison,
  title    = "Comparing Physician and Artificial Intelligence Chatbot Responses
              to Patient Questions Posted to a Public Social Media Forum",
  author   = "Ayers, John W and Poliak, Adam and Dredze, Mark and Leas, Eric C
              and Zhu, Zechariah and Kelley, Jessica B and Faix, Dennis J and
              Goodman, Aaron M and Longhurst, Christopher A and Hogarth,
              Michael and Smith, Davey M",
  journal  = "JAMA Intern Med",
  volume   =  183,
  number   =  6,
  pages    = "589--596",
  month    =  jun,
  year     =  2023,
  address  = "United States",
  language = "en"
}

@article{braun2006thematicanalysis,
author = {Virginia Braun and Victoria Clarke},
title = {Using thematic analysis in psychology},
journal = {Qualitative Research in Psychology},
volume = {3},
number = {2},
pages = {77--101},
year = {2006},
publisher = {Routledge},
doi = {10.1191/1478088706qp063oa},
URL = {         https://www.tandfonline.com/doi/abs/10.1191/1478088706qp063oa
},
eprint = {         https://www.tandfonline.com/doi/pdf/10.1191/1478088706qp063oa
}
}

@inproceedings{wang2020wechatpatientcarechina,
author = {Wang, Ding and Kale, Santosh D. and O'Neill, Jacki},
title = {Please Call the Specialism: Using WeChat to Support Patient Care in China},
year = {2020},
isbn = {9781450367080},
publisher = {ACM},
address = {New York, NY, USA},
url = {https://doi.org/10.1145/3313831.3376274},
doi = {10.1145/3313831.3376274},
booktitle = {Proceedings of the 2020 CHI Conference on Human Factors in Computing Systems},
pages = {1–13},
numpages = {13},
location = {Honolulu, HI, USA},
series = {CHI '20}
}

@article{au2023ainotready,
  title={AI chatbots not yet ready for clinical use},
  author={Au Yeung, Joshua and Kraljevic, Zeljko and Luintel, Akish and Balston, Alfred and Idowu, Esther and Dobson, Richard J and Teo, James T},
  journal={Frontiers in Digital Health},
  volume={5},
  pages={60},
  year={2023},
  publisher={Frontiers}
}

@article{mcghee2020cataractcommon,
  title={A perspective of contemporary cataract surgery: the most common surgical procedure in the world},
  author={McGhee, Charles NJ and Zhang, Jie and Patel, Dipika V},
  journal={Journal of the Royal Society of New Zealand},
  volume={50},
  number={2},
  pages={245--262},
  year={2020},
  publisher={Taylor \& Francis}
}

@Article{thompson2024leafletscataract,
author={Thompson, Polly
and Thornton, Richard
and Ramsden, Conor M.},
title={Assessing chatbots ability to produce leaflets on cataract surgery: Bing AI, chatGPT 3.5, chatGPT 4o, ChatSonic, Google Bard, Perplexity, and Pi},
journal={Journal of Cataract {\&} Refractive Surgery},
year={2025},
volume={51},
pages={371-375},
number={5},
issn={0886-3350},
url={https://journals.lww.com/jcrs/fulltext/2025/05000/assessing_chatbots_ability_to_produce_leaflets_on.5.aspx}
}

@misc{wadhwa2025designingculturesocialnorms,
      title={Designing with Culture: How Social Norms Shape Trust and Preference in Health Chatbots}, 
      author={Arpita Wadhwa and Aditya Vashistha and Mohit Jain},
      year={2025},
      eprint={2509.15575},
      archivePrefix={arXiv},
      primaryClass={cs.HC},
      url={https://arxiv.org/abs/2509.15575}, 
}

@article{sachdeva2024learningslargescaledeploymentllmpowered,
author = {Bhuvan Sachdeva and Pragnya Ramjee and Rahul Sharma and Mithun Thulasidas and Geeta Fulari and Kaushik Murali and Mohit Jain},
title ={Utility of an LLM-Powered Experts-in-the-Loop Chatbot for Pre- and Post-operative Care of Cataract Surgery Patients},
journal = {European Journal of Ophthalmology},
year = {2025},
}

@inproceedings{inform-bot-design-dis18,
author = {Jain, Mohit and Kumar, Pratyush and Kota, Ramachandra and Patel, Shwetak N.},
title = {Evaluating and Informing the Design of Chatbots},
year = {2018},
isbn = {9781450351980},
publisher = {ACM},
address = {New York, NY, USA},
url = {https://doi.org/10.1145/3196709.3196735},
doi = {10.1145/3196709.3196735},
booktitle = {Proceedings of the 2018 Designing Interactive Systems Conference},
pages = {895–906},
numpages = {12},
location = {Hong Kong, China},
series = {DIS '18}
}

@inproceedings{ramjee2025ashabot,
author = {Ramjee, Pragnya and Chhokar, Mehak and Sachdeva, Bhuvan and Meena, Mahendra and Abdullah, Hamid and Vashistha, Aditya and Nagar, Ruchit and Jain, Mohit},
title = {ASHABot: An LLM-Powered Chatbot to Support the Informational Needs of Community Health Workers},
year = {2025},
isbn = {9798400713941},
publisher = {ACM},
address = {New York, NY, USA},
url = {https://doi.org/10.1145/3706598.3713680},
doi = {10.1145/3706598.3713680},
booktitle = {Proc of the 2025 CHI Conference on Human Factors in Computing Systems},
articleno = {698},
numpages = {22},
series = {CHI '25}
}

@inproceedings{reza2024abscribehumanaicowriting,
author = {Reza, Mohi and Laundry, Nathan M and Musabirov, Ilya and Dushniku, Peter and Yu, Zhi Yuan “Michael” and Mittal, Kashish and Grossman, Tovi and Liut, Michael and Kuzminykh, Anastasia and Williams, Joseph Jay},
title = {ABScribe: Rapid Exploration \& Organization of Multiple Writing Variations in Human-AI Co-Writing Tasks using Large Language Models},
year = {2024},
isbn = {9798400703300},
publisher = {ACM},
address = {New York, NY, USA},
url = {https://doi.org/10.1145/3613904.3641899},
doi = {10.1145/3613904.3641899},
booktitle = {Proc of the 2024 CHI Conference on Human Factors in Computing Systems},
articleno = {1042},
numpages = {18},
location = {Honolulu, HI, USA},
series = {CHI '24}
}

@incollection{hancock1988nasatlx,
title = {Development of NASA-TLX (Task Load Index): Results of Empirical and Theoretical Research},
series = {Advances in Psychology},
publisher = {North-Holland},
volume = {52},
pages = {139-183},
year = {1988},
booktitle = {Human Mental Workload},
issn = {0166-4115},
doi = {https://doi.org/10.1016/S0166-4115(08)62386-9},
url = {https://www.sciencedirect.com/science/article/pii/S0166411508623869},
author = {Sandra G. Hart and Lowell E. Staveland},
address={USA},
}

@inproceedings{chen2025healthtechforimmigrants,
author = {Chen, Zhanming and Ghaju, Alisha and Hang, May and Maestre, Juan Fernando and Shin, Ji Youn},
title = {Designing Health Technologies for Immigrant Communities: Exploring Healthcare Providers' Communication Strategies with Patients},
year = {2025},
isbn = {9798400713941},
publisher = {Association for Computing Machinery},
address = {New York, NY, USA},
url = {https://doi.org/10.1145/3706598.3713782},
doi = {10.1145/3706598.3713782},
booktitle = {Proceedings of the 2025 CHI Conference on Human Factors in Computing Systems},
articleno = {1045},
numpages = {19},
series = {CHI '25}
}

@inproceedings{liu2025humanizedllms,
author = {Liu, Dingdong and Zhang, Yujing and Zhao, Bolin and Ma, Shuai and Shi, Chuhan and Ma, Xiaojuan},
title = {Scaffolded Turns and Logical Conversations: Designing Humanized LLM-Powered Conversational Agents for Hospital Admission Interviews},
year = {2025},
isbn = {9798400713941},
publisher = {Association for Computing Machinery},
address = {New York, NY, USA},
url = {https://doi.org/10.1145/3706598.3714196},
doi = {10.1145/3706598.3714196},
booktitle = {Proceedings of the 2025 CHI Conference on Human Factors in Computing Systems},
articleno = {643},
numpages = {23},
series = {CHI '25}
}

@inproceedings{joshi2025userpereferencesaihealth,
author = {Joshi, Rutuja and Lee, Yu-Jou and Bengler, Klaus},
title = {User Preferences in Conversational AI for Healthcare: Insights from an Interview Study},
year = {2025},
isbn = {9798400715273},
publisher = {Association for Computing Machinery},
address = {New York, NY, USA},
url = {https://doi.org/10.1145/3719160.3736631},
doi = {10.1145/3719160.3736631},
booktitle = {Proceedings of the 7th ACM Conference on Conversational User Interfaces},
articleno = {67},
numpages = {13},
series = {CUI '25}
}

@article{amershi2014powertothepeople,
author = {Amershi, Saleema and Cakmak, Maya and Knox, W. Bradley and Kulesza, Todd},
title = {Power to the People: The Role of Humans in Interactive Machine Learning},
year = {2014},
issue_date = {Winter 2014},
publisher = {John Wiley \& Sons, Inc.},
address = {USA},
volume = {35},
number = {4},
issn = {0738-4602},
url = {https://doi.org/10.1609/aimag.v35i4.2513},
doi = {10.1609/aimag.v35i4.2513},
journal = {AI Mag.},
month = dec,
pages = {105–120},
numpages = {16}
}

@inproceedings{horvitz1999mixedinitiative,
author = {Horvitz, Eric},
title = {Principles of mixed-initiative user interfaces},
year = {1999},
isbn = {0201485591},
publisher = {Association for Computing Machinery},
address = {New York, NY, USA},
url = {https://doi.org/10.1145/302979.303030},
doi = {10.1145/302979.303030},
booktitle = {Proceedings of the SIGCHI Conference on Human Factors in Computing Systems},
pages = {159–166},
numpages = {8},
location = {Pittsburgh, Pennsylvania, USA},
series = {CHI '99}
}

\newpage
\section*{NeurIPS Paper Checklist}

\begin{enumerate}

\item {\bf Claims}
    \item[] Question: Do the main claims made in the abstract and introduction accurately reflect the paper's contributions and scope?
    \item[] Answer: \answerYes{}
    \item[] Justification: The claims in the abstract and introduction reflect the scope and contributions of our research.
    
\item {\bf Limitations}
    \item[] Question: Does the paper discuss the limitations of the work performed by the authors?
    \item[] Answer: \answerYes{}
    \item[] Justification: Please see Section \ref{limitations}.
    
\item {\bf Theory assumptions and proofs}
    \item[] Question: For each theoretical result, does the paper provide the full set of assumptions and a complete (and correct) proof?
    \item[] Answer: \answerNA{}
    \item[] Justification: We do not produce theory.
    
\item {\bf Experimental result reproducibility}
    \item[] Question: Does the paper fully disclose all the information needed to reproduce the main experimental results of the paper to the extent that it affects the main claims and/or conclusions of the paper (regardless of whether the code and data are provided or not)?
    \item[] Answer: \answerYes{}
    \item[] Justification: Please see Section \ref{methods}.

\item {\bf Open access to data and code}
    \item[] Question: Does the paper provide open access to the data and code, with sufficient instructions to faithfully reproduce the main experimental results, as described in supplemental material?
    \item[] Answer: \answerNA{}
    \item[] Justification: The paper does not include experiments requiring code.

\item {\bf Experimental setting/details}
    \item[] Question: Does the paper specify all the training and test details (e.g., data splits, hyperparameters, how they were chosen, type of optimizer, etc.) necessary to understand the results?
    \item[] Answer: \answerNA{}.
    \item[] Justification: Our experiment doesn't involve training an AI model and hence no data split or hyperparameter tuning reported. 
    
\item {\bf Experiment statistical significance}
    \item[] Question: Does the paper report error bars suitably and correctly defined or other appropriate information about the statistical significance of the experiments?
    \item[] Answer: \answerYes{}
    \item[] Justification: Results are accompanied by statistical significance tests and associated explanations.

\item {\bf Experiments compute resources}
    \item[] Question: For each experiment, does the paper provide sufficient information on the computer resources (type of compute workers, memory, time of execution) needed to reproduce the experiments?
    \item[] Answer: \answerNA{}
    \item[] Justification: Our experiment doesn't involve training an AI model.
    
\item {\bf Code of ethics}
    \item[] Question: Does the research conducted in the paper conform, in every respect, with the NeurIPS Code of Ethics \url{https://neurips.cc/public/EthicsGuidelines}?
    \item[] Answer: \answerYes{} 
    \item[] Justification: We adhere completely to the NeurIPS Code of Ethics.
    
\item {\bf Broader impacts}
    \item[] Question: Does the paper discuss both potential positive societal impacts and negative societal impacts of the work performed?
    \item[] Answer: \answerYes{}
    \item[] Justification: Please see Section \ref{discussion}.
    
\item {\bf Safeguards}
    \item[] Question: Does the paper describe safeguards that have been put in place for responsible release of data or models that have a high risk for misuse (e.g., pretrained language models, image generators, or scraped datasets)?
    \item[] Answer: \answerNA{}
    \item[] Justification: Our experiment doesn't involve training an AI model.

\item {\bf Licenses for existing assets}
    \item[] Question: Are the creators or original owners of assets (e.g., code, data, models), used in the paper, properly credited and are the license and terms of use explicitly mentioned and properly respected?
    \item[] Answer: \answerYes{}
    \item[] Justification: Please see Section \ref{methods}.

\item {\bf New assets}
    \item[] Question: Are new assets introduced in the paper well documented and is the documentation provided alongside the assets?
    \item[] Answer: \answerNA{}
    \item[] Justification: Our experiment doesn't involve new AI model/data assets.
    
\item {\bf Crowdsourcing and research with human subjects}
    \item[] Question: For crowdsourcing experiments and research with human subjects, does the paper include the full text of instructions given to participants and screenshots, if applicable, as well as details about compensation (if any)? 
    \item[] Answer: \answerYes{} 
    \item[] Justification: Please see Section \ref{methods} and Figure \ref{fig:docedit_portal}.
    
\item {\bf Institutional review board (IRB) approvals or equivalent for research with human subjects}
    \item[] Question: Does the paper describe potential risks incurred by study participants, whether such risks were disclosed to the subjects, and whether Institutional Review Board (IRB) approvals (or an equivalent approval/review based on the requirements of your country or institution) were obtained?
    \item[] Answer: \answerYes{}
    \item[] Justification: IRB approval was obtained, and risk are highlighted and mitigated, please see Section \ref{methods}.
    
\item {\bf Declaration of LLM usage}
    \item[] Question: Does the paper describe the usage of LLMs if it is an important, original, or non-standard component of the core methods in this research? Note that if the LLM is used only for writing, editing, or formatting purposes and does not impact the core methodology, scientific rigorousness, or originality of the research, declaration is not required.
    \item[] Answer: \answerYes{}
    \item[] Justification: Please see Section \ref{methods}.
    
\end{enumerate}

\end{document}